\documentclass[useAMS]{mn2e}
\usepackage{times}

\input{epsf}

\title[Black hole accretion discs: reality confronts theory]
{Black hole accretion discs: reality confronts theory}

\author[M. Gierli\'nski, C. Done]
{Marek~Gierli\'nski$^{1,2}$ and Chris Done$^1$\\
$^1$Department of Physics, University of Durham, South Road, Durham 
DH1 3LE, UK\\ 
$^2$Obserwatorium Astronomiczne Uniwersytetu Jagiello{\'n}skiego, 
30-244 Krak{\'o}w, Orla 171, Poland}

\date{Submitted to MNRAS}
\pagerange{\pageref{firstpage}--\pageref{lastpage}}
\pubyear{2002}

\begin{document}

\topmargin = -0.5cm

\maketitle

\label{firstpage}

\begin{abstract}

Disc spectra from highly luminous black hole binaries are observed to
be rather simple, despite theoretical predictions to the contrary.  We
collate the disc-dominated spectra from multiple observations of 10
separate sources, and show that these overwhelmingly follow a
multi-temperature disc blackbody spectrum with luminosity $\propto
T^4$, and that such discs are stable. These results are in conflict
with standard Shakura-Sunyaev $\alpha$-discs predictions, including
proposed modifications such as additional energy loss in a jet, wind
or corona. Accretion disc spectra can be useful, reliable guides to
the disc structure, and should be used to test the next generation of
accretion disc models, in which the disc viscosity is calculated
self-consistently from the magnetically generated turbulent dynamo.

\end{abstract}

\begin{keywords}
  accretion, accretion discs -- X-rays: binaries
\end{keywords}


\section{Introduction}

Accretion onto a compact object is believed to be the source of power
in X-ray binaries (see e.g.~Tanaka \& Lewin 1995). At accretion rates
in excess of a few percent of Eddington rate most of the black hole
binaries dwell in the so-called soft, or high, X-ray spectral state.
Their X-ray spectra are then dominated by a soft thermal component, of
typical temperature $\la$ 1~keV. The standard model of accretion
(Shakura \& Sunyaev 1973, hereafter SS73) explains these spectra as
emission from an optically thick and geometrically thin disc.  This
can form a stable emitting structure down to the minimum stable orbit
(which depends on the black hole spin),
so its spectrum contains the imprint of strong special and
general relativistic effects (e.g.\ Cunningham 1975; Ebisawa, Mitsuda
\& Hanawa 1991). 

This all makes disc spectra an exciting potential probe of the 
dramatically curved space-time in the vicinity of the black hole. However, 
there are many problems in interpreting them. Firstly, the 
quasi-thermal component which is identified as the disc emission is 
generally accompanied by a power-law tail out to much higher energies 
(e.g.\ Tanaka \& Lewin 1995). This tail is well fit by Comptonization 
of seed photons from the disc, requiring that some of the energy is 
{\em not} thermalized in the disc but is instead dissipated in an 
optically thin environment, e.g.\ a corona above the disc. The disc 
structure here {\em must} be different to that assumed by disc 
equations where all the energy is dissipated in the optically thick 
material. Secondly, these coronal hard X-rays can illuminate the 
disc, again changing its structure (e.g.\ Nayakshin, Kazanas \& 
Kallman 2000a). Thirdly, even without the coronal emission, the disc 
spectrum at a given radius is not a simple blackbody as the 
absorption opacity is small at the high temperatures expected for 
X-ray binary discs (SS73). The amount of deviation from blackbody 
(termed a colour temperature correction) depends on details of the 
disc vertical structure, but the net result is to increase the 
observed peak `temperature' of the disc (SS73)

A pessimistic response to this complexity is simply to abandon 
any information the disc spectra might contain as being too 
difficult to disentangle.  However, some black holes can show 
spectra which are dominated by the disc component, with very 
little hard X-ray emission. These offer a way of circumventing 
the first two problems described above, so these spectra can 
equally be viewed as giving {\em observational} constraints on 
the disc structure, black hole mass and spin (see e.g. Ebisawa 
et al.~1991; 1994; Kubota, Makishima \& Ebisawa 2001; Kubota \& 
Makishima 2003)

Further motivation for this more optimistic approach is that the 
disc stability is also a sensitive indicator of how the 
gravitational energy is dissipated within the disc. Standard 
discs ($\alpha$-discs) in which the heating is proportional to 
the total (gas plus radiation) pressure, are viscously (Lightman 
\& Eardley 1974) and thermally (Shakura \& Sunyaev 1976) 
unstable when radiation pressure dominates. In this regime the 
disc undergoes limit-cycle variability on time-scales of a few 
hundred seconds between a low accretion rate, gas pressure 
dominated state and a high accretion rate, advectively cooled 
slim disc (Honma, Matsumoto \& Kato 1991; Szuszkiewicz \& Miller 
1997, 1998; Zampieri, Turolla \& Szuszkiewicz 2001).

Here we compile all the available {\it Rossi X-ray Timing 
Explorer\/} {\it RXTE\/} observations of several black hole 
binaries which show disc-dominated spectra to see what 
information can be reliably extracted from the data.  The 
observed disc spectra are surprisingly simple. They can be well 
modelled by a multi-colour disc blackbody, and individual 
objects which span a large range in disc luminosity, $L_{\rm 
disc}$, generally have an observed disc temperature $\propto 
L_{\rm disc}^{1/4}$. We use the sample as a whole to set 
constraints on the colour temperature correction and limits on 
how this changes as a function of luminosity, and show that such 
spectra {\em can} be used to derive black hole spin if the 
binary system parameters are well known.  We use the data on the 
colour temperature correction, together with the observed 
stability of the light curves, to constrain the disc structure. 
We show it must be somewhat different to that predicted by the 
standard $\alpha$-disc viscous heating prescription, 
irrespective of phenomenological modifications such as 
winds/jets/coronae. The data are also probably inconsistent with 
an alternative viscosity prescription in which the heating is 
proportional to gas pressure alone (Stella \& Rosner 1984).

While no current models can fully explain the observations, we stress
that disc spectra {\em can be} a simple, useful and reliable probe of the
accretion mechanisms. These tests will be especially important 
for the next generation of models which will calculate the heating 
{\it ab initio} from the MHD dynamo (Balbus \& Hawley 1991) rather 
than using the {\it ad hoc} viscosity prescriptions described above.

\section{Disc temperature and luminosity}

An optically thick, geometrically thin accretion disc around a 
non-rotating black hole has a very robust predicted spectrum. It does 
not depend on the details of the viscous heating if the gravitational 
energy released at a given radius $R$ is emitted locally as a 
blackbody of temperature $T_{\rm eff}(R)$.  The total spectrum of the 
disc is the sum of these blackbodies (generally termed a multicolour 
disc blackbody), with maximum temperature $T_{\rm eff,max}$ occurring 
close to the minimum stable orbit. The luminosity of the disc is 
$L_{\rm disc} \propto T_{\rm eff,max}^4$, and this completely 
determines the disc spectrum. 

However, a given radius in the disc need not emit as a true 
blackbody. The absorption opacity at the high temperatures 
characteristic of X-ray binary discs is small, so scattering is 
important. In this case the disc spectrum is either a modified 
blackbody (with detailed shape depending on the vertical density and 
temperature structure of the disc) or Comptonized into a Wien peak 
(SS73).  When these are taken into account, the emergent disc 
spectrum can still be approximated by a multicolour disc blackbody 
but of maximum {\em colour\/} temperature, $T_{\rm max} \equiv f_{\rm 
col} T_{\rm eff,max}$ (Shimura \& Takahara 1995; Merloni et 
al.~2000), where $f_{\rm col}$ is called the colour temperature 
correction, or spectral hardening factor, since it is always greater 
then unity. 

To find the numerical coefficient linking luminosity and temperature, 
we assume the disc extends down to the minimum stable orbit at $6R_g$
in a pseudo-Newtonian potential (Paczy{\'n}ski \& Wiita 1980), 
\begin{equation}\Phi(R) = - {GM \over R - 2R_g},\end{equation} where 
$G$ is the gravitational constant, $M$ is the black hole mass and $R_g
\equiv GM/c^2$ is the gravitational radius. This is a simple and
accurate approximation of the gravitational potential around a
non-rotating black hole. The maximum disc temperature is produced at
$\approx 9.5R_g$ with the stress-free inner boundary condition
(e.g.~Novikov \& Thorne 1973).  Appendix A of Gierli{\'n}ski et
al.~(1999) shows that the integrated disc luminosity is related to the
maximum observed colour temperature, $T_{\rm max}$, as
\begin{equation} L_{\rm disc} = {\pi \sigma G^2 M^2 T_{\rm max}^4
\over 6 c_0^4 f_{\rm col}^4 c^4}. \end{equation} Here $\sigma$ is the
Stefan-Boltzmann constant and $c_0 \approx 0.1067$. The above formula
can be expressed in terms of Eddington luminosity,
$L_{\rm Edd} = 1.48 \times 10^{38} (M / $M$_\odot)$ erg s$^{-1}$, as
\begin{equation} {L_{\rm disc} \over L_{\rm Edd}} \approx 0.583 
\left( 1.8 \over f_{\rm col} \right)^{4} \left( M \over 10 M_\odot 
\right) \left( kT_{\rm max} \over \rm{1~keV} \right)^4, 
\label{eq:lt4}\end{equation} where $k$ is the Boltzmann constant. 

There is considerable speculation as to whether the inner stress-free 
boundary condition is accurate. The viscosity is now known to be 
derived from an MHD dynamo, and magnetic reconnection can give 
continuous stress across the last stable orbit (Agol \& Krolik 2000). 
Ironically, in this case the simple {\sc diskbb} model (Mitsuda et 
al.~1984; in the {\sc xspec} spectral fitting package) is more 
appropriate as it does not incorporate the stress-free boundary 
condition (though it also has a Newtonian potential rather than 
pseudo-Newtonian).  In any case, the shape of the {\sc diskbb} 
spectrum is extremely close to that derived from the 
pseudo-Newtonian, stress-free disc model ({\sc diskpn} in {\sc 
xspec}) described above, requiring only a 4 per cent temperature 
shift to match the spectrum at all energies above $\sim 0.5 kT_{max}$ 
(Gierli{\'n}ski et al.~1999). Thus we use the {\sc diskbb} model with 
an additional temperature correction factor of $\xi \approx 1.04$.

The emerging disc emission is then affected by the relativistic 
effects in strong gravitational potential of the black hole. 
Following Cunningham (1975) and Zhang, Cui \& Chen (1997) we apply 
corrections $g$ and $f_{\rm GR}$ for the observed flux and 
temperature, respectively: \begin{equation} F' = g(i, a_*) {L_{\rm 
disc} \over 2 \pi D^2}\label{eq:fgr}\end{equation} and 
\begin{equation}T'_{\rm max} = f_{\rm GR}(i, a_*) \xi T_{\rm 
max}.\label{eq:tgr}\end{equation} where $i$ is the inclination angle, 
and $D$ is the distance to the source. The dimensionless spin is 
defined as $a_* \equiv Jc/GM^2$, where $J$ is the angular momentum of 
the black hole. The primed quantities denote values in the observer's 
frame. We use the values of $g$ and $f_{\rm GR}$ from table 1 in 
Zhang et al.~(1997).



\begin{figure*}
\begin{center}
\leavevmode
\epsfxsize=13cm \epsfbox{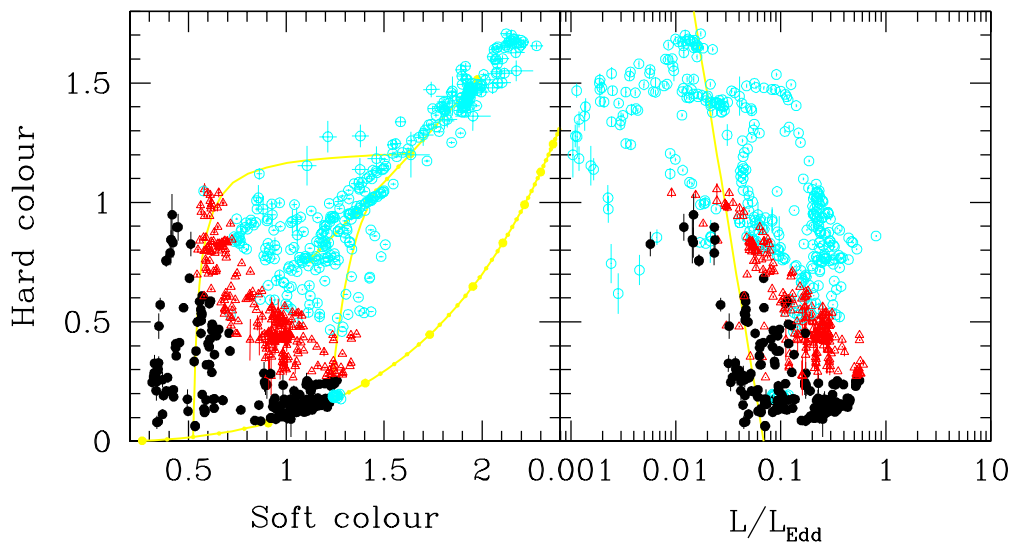}
\end{center}

\caption{The colour-colour (left) and colour-luminosity (right) 
diagram of the sources listed in Table \ref{tab:sources}. See 
DG03 for detailed description of these diagrams. Different 
symbols correspond to different fraction of the total luminosity 
released in the Comptonized tail above the disc spectrum: $\le$ 
5 per cent (black filled circles), 5--15 per cent (red open 
triangles) and $>$ 15 per cent (cyan open circles). Only the 
data with disc-dominated spectra ($\le$ 15 per cent of 
Comptonized emission) are analysed in this paper. See electronic 
edition of the journal for the colour version of this figure.}

\label{fig:colum}
\end{figure*}


\section{Sample Selection}
\label{sec:sample}

We want to select disc-dominated spectra, where contribution from the 
Comptonized tail is as small as possible. We use the black hole 
sample of Done \& Gierli{\'n}ski~(2003, hereafter DG03) as our basis. 
This consists of all the well observed {\it RXTE\/} black holes with 
low absorption which showed state transitions, i.e. these include 
soft state spectra.  We extend the sample to include XTE~J1650--500 
and two black hole systems which were always in the soft state when 
observed with {\it RXTE\/} (LMC~X-1 and GRS~1739--278).  We also 
include data from the {\it Ginga\/} black hole transient GS~1124--68 
(a.k.a. Nova Muscae), as this showed a soft (disc dominated) spectrum 
over much of its outburst (Ebisawa et al.~1994).

To select the disc-dominated spectra from the whole sample we fit the 
data by the model consisting of the disc emission and its thermal 
Comptonization. We describe the model and fitting procedure in the 
next section. Ideally we would require that there was {\em no} 
Comptonized tail present. However, this sort of ultrasoft spectrum is 
not very common, so in order to have a larger sample we allow up to 
15 per cent of the total bolometric luminosity to be present in the 
tail. This is somewhat model dependent, but the precise value of the 
coronal fraction is not important as long as it is small enough to 
have a negligible effect on the disc structure.

Fig. \ref{fig:colum} shows the complete intrinsic colour-colour and 
colour-luminosity diagram as in DG03 for our observed sources.  The 
lower curving line in the background shows the colours expected for a 
pure {\sc diskbb} spectrum with temperature increasing along the line 
from 0.5~keV (lower left) to 4.5~keV (top right); see also fig.~1 in 
DG03. This line forms a lower bound on the disc-dominated spectra. 
The Comptonized fraction increases with increasing both colours, 
roughly along the diagonal. The disc-dominated data selected for 
further analysis are marked by black filled circles and red open 
triangles, to denote spectra where the Compton tail fractional 
luminosity is between 0--5 and 5--15 per cent, respectively. Typical 
X-ray spectra for the selected data are shown in panels $a$, $g$, 
$h$, $i$ and $j$ of fig. 3 in DG03. 

The limit of 15 per cent of flux in the Compton tail excludes all data
from XTE~J1118+480, a source well observed by {\it RXTE} (e.g.
McClintock et al.~2001), as it has been seen only in the hard state.
More surprisingly perhaps it also excludes Cyg X-1 as although this
makes a transition to the soft state, its X-ray spectra always include
a substantial fraction of Comptonized emission (see e.g.
Gierli{\'n}ski et al.~1999; Frontera et al.~2001; DG03). The
interesting source GRS~1915+105 is also excluded as again the spectra
generally show substantial Comptonized emission as well as having high
(and probably variable) absorption which complicates the spectral
fitting. This gives a sample of 10 black hole systems, as listed in
Table \ref{tab:sources}. In this table we also list the mass,
distance, inclination and absorption column used in the analysis,
together with their uncertainties. There are two sets of distance estimates
quoted for GRO~J1655--40, as the recently updated value (Mirabel et
al.~2002) is significantly different to the previous determination.

\begin{figure}
\begin{center}
\leavevmode
\epsfxsize=8.5cm \epsfbox{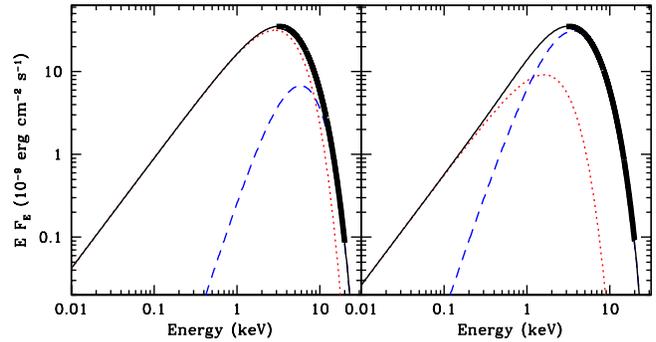}
\end{center}

\caption{Unabsorbed model components fitted to the spectrum of 
GRO~J1655--40 corresponding to a point in the colour-colour 
diagram (Fig.~\ref{fig:colum}, left panel) of colours (1.25, 
0.20). The softer component (dotted line) represents the disc 
emission, the harder one (dashed line) -- its Comptonization. 
The heavy line depicts the observed data. The two different 
models presented here give a similar fit to the data. However, 
the disc-dominated model (on the left) requires an unphysically 
small inner disc radius.}

\label{fig:warning} 
\end{figure}

The intrinsic colour-colour diagram (Fig.~\ref{fig:colum}) shows an 
almost one-to-one match between colour and the fraction of 
Comptonization inferred from the spectrum. There is, however, one 
exception -- a cluster of points from GRO~J1655--40 around colours 
(1.25, 0.20). They lie almost on the disc blackbody line and their 
spectra {\em can} be fitted by the disc blackbody model with 
temperature of $\sim$ 1.4~keV. However, as shown by Kubota et 
al.~(2001), these spectra can be also interpreted as emission from 
the disc of lower temperature plus additional Comptonization in an 
optically thick plasma. The details of spectral modelling differ 
slightly between Kubota et al.~(2001) and this work, in particular we 
use a Comptonization model rather than a power law to characterize 
the tail. However, the general result is the same: these particular 
spectra can be fitted either by a dominating disc, or by a much weaker 
disc plus strong Comptonization. In Fig.~\ref{fig:warning} we show the 
unabsorbed model components of both fits. They both give a similar 
$\chi^2$ but the disc-dominated model (left panel in 
Fig.~\ref{fig:warning}) yields an unphysically small inner disc 
radius (see also Kubota et al.~2001). Therefore, we select the 
strongly Comptonized model (right panel in Fig.~\ref{fig:warning}) as 
the more physically realistic. Since these spectra are then dominated 
by the Comptonized emission rather than by the disc they are excluded 
from further analysis.

\begin{table*}
\begin{tabular}{lccccc}
\hline 
Source Name & $M$ (M$_\odot$) & $D$ (kpc) & $i$ (deg) & $N_H$ ($10^{22}$ cm$^{-2}$)\\
\hline
LMC X-3 & 7 (5--11)$^a$& 52 (51.4--52.6)$^b$ & 60 (50--70)$^c$ & 0.07$^d$\\
LMC X-1 & 10 (4--12.5)$^e$ & 52 (51.4--52.6)$^b$ & 45 (24--64)$^e$ & 0.5$^f$\\
GS 1124--68 & 7 (6.4--7.6)$^g$ & 2.8 (2.8--4)$^h$ & 54 (52.5--55.5)$^g$ & 0.16$^i$\\
XTE J1550--564 & 10 (9.7--11.6)$^j$ & 5.3 (2.8--7.6)$^j$ & 72 (70.8--75.4)$^j$ & 0.65$^k$\\
XTE J1650--500 & [10] & 4 (2--6)$^l$ & 30 ($<$40)$^m$ & 0.78$^n$\\
GRO J1655--40  & 7 (6.8--7.2)$^o$ & 3.2$\pm$0.2$^p$ or 0.9$\pm$0.1$^q$ & 70 (64--71)$^r$ & 0.8$^f$\\
GX 339--4 & 6 (2.5--10)$^s$ & 4 (2.6--5)$^t$ & 40 (20--60)$^s$ & 0.6$^t$\\
GRS 1739--278 & [10] & 8.5 (6--11)$^u$ & [60] & 2$^u$\\
XTE J1859+226 & [10] & 7.6 (4.6--8)$^v$ & [60] & 0.8$^w$\\
XTE J2012+381 & [10] & [8.5] & [60] & 1.3$^x$\\
\hline
\end{tabular}

\caption{The list of the sources used in this paper, together with 
assumed mass ($M$), distance ($D$), disc inclination ($i$) and 
absorption column ($N_H$) and their uncertainties. The numbers in 
square brackets correspond to assumed values where the constraints 
are not known. The numbered references are as follows:
[a] Soria et al.~2001
[b] di Benedetto 1997
[c] Cowley et al.~1983
[d] Haardt et al.~2001
[e] Hutchings et al.~1987
[f] Gierli{\'n}ski, Macio{\l}ek-Nied{\'z}wiecki \& Ebisawa 2001
[g] Gelino, Harrison \& McNamara 2001
[h] Shahbaz, Naylor \& Charles 1997
[i] {\.Z}ycki, Done \& Smith 1998
[j] Orosz et al.~2002
[k] Gierli{\'n}ski \& Done 2003
[l] Tomsick et al.~2003
[m] S{\'a}nchez-Fern{\'a}ndez et al.~2002
[n] Miller et al.~2002
[o] Shahbaz et al.~1999
[p] Hjellming \& Rupen 1995
[q] Mirabel et al.~2002
[r] van der Hooft et al.~1998
[s] Cowley et al.~2002
[t] Zdziarski et al.~1998
[u] Greiner, Dennerl \& Predehl 1996
[v] Hynes et al.~2002
[w] dal Fiume et al.~1999
[x] Campana et al.~2002
}

\label{tab:sources} \end{table*}


\section{Spectral modelling}
\label{sec:fitting}

For the {\it RXTE\/} sources we use the same data selection as DG03, 
i.e. extract one Proportional Counter Array (PCA) spectrum per one 
pointing (designated by a unique observation ID), typically giving 
exposures of a few kiloseconds. We use data from detectors 0, 2 and 
3, top layer only, except for GRO~J1655--40 and XTE~J1739--278, where 
detector 3 was off most of the time, so we used detectors 0 and 2 
only. We fit these spectra in 3--20 keV band, after adding 1 per cent 
systematic error in each energy channel.  For the {\it Ginga} 
observations of GS~1124--68 we use the spectra extracted by 
{\.Z}ycki et al.~(1998) and fit these over the 2--20 keV 
bandpass with 0.5 per cent systematic error.

We follow DG03 and fit the data with a model consisting of 
multicolour disc blackbody ({\sc diskbb} in {\sc xspec}) and its 
thermal Comptonization (Zdziarski, Johnson \& Magdziarz 1996). We add 
a Gaussian line and smeared edge in the iron K$\alpha$ complex to 
approximately account for the effects of Compton reflection. This 
adequately fits {\em all} the data from all black hole (and neutron 
star) spectral states (DG03).  We stress that though the model is 
highly simplified, it is appropriate for our purpose. We are not 
interested here in detailed spectral modelling of Comptonized 
emission (which is probably non-thermal rather than thermal in the 
soft state: Gierli{\'n}ski et al.~1999) nor in its reflection. Since 
we select the only disc-dominated spectra, the details of the 
Comptonization/reflection modelling  affect the disc results only 
very weakly. On the other hand, our Comptonization model includes a 
proper low-energy cutoff around seed photon energy, which {\em is} 
important. An extrapolation of a power law to low energies predicts 
much more flux in the tail at the seed photon energy than that given 
by a Comptonization model. This decreases the inferred disc flux, 
especially for the steep tails associated with the soft states (see 
e.g. Done, {\.Z}ycki \& Smith 2002).

The highest disc temperature in the sample is $\approx$ 1.3 keV. The 
maximum power of the {\sc diskbb} model is emitted at around $2.4 
kT_{\rm max}$. The useful observing bandpass of the PCA extends down 
to about 3 keV, below which the detector's sensitivity rapidly drops 
and systematic uncertainties become important. This means that the 
PCA can only seldom observe the peak of the disc emission, though it 
can still see a large fraction of its Wien tail. For example, a 
multicolour disc with $kT_{\rm max} = 0.5$ keV has an amplitude of 
$\sim$ 30 per cent of maximum at 3 keV in a $\nu F(\nu)$ spectrum. 
Therefore it is feasible for the PCA to effectively measure the 
observed disc temperatures down to $\sim$ 0.5 keV. While fitting the 
spectra we limit $kT_{\rm max} \ge 0.4$ keV (or 0.3~keV for the {\it 
Ginga} data as these have bandpass down to 2~keV) and reject 
observations for which the measured $kT_{\rm max}$ are consistent 
with this lower limit (within statistical errors) as potentially 
unconstrained.

The temperature of the Comptonized component is difficult to 
constrain from the PCA data alone, particularly in the soft state, 
where the high-energy tail is weak. Leaving this parameter free makes 
deriving uncertainties of the disc parameters difficult for most of 
the soft-state spectra. Therefore, we fix it at 50~keV, which is well 
above the high energy limit of the PCA data of 20~keV. This makes no 
difference to the derived disc parameters from the disc-dominated 
spectra.

This model fits the data very well indeed. We fit all the 
observations in {\sc xspec}, obtaining reduced $\chi^2/\nu < 1.5$ 
(except for one GRO~J1655--40 observation with strong absorption 
lines, see Ueda et al.~1998). As a result we get the disc 
temperature, $T'_{\rm max}$, and the unabsorbed, bolometric disc flux 
$F'$. From this we calculate the luminosity in Eddington units, 
$L_{\rm disc}/L_{\rm Edd}$, and temperature corrected for 
relativistic effects, $T_{\rm max}$, using formulas (\ref{eq:fgr}) 
and (\ref{eq:tgr}), assuming a non-rotating black hole, $a_* = 0$ and 
mass and distance for each source from Table \ref{tab:sources}.


\begin{figure*}
\begin{center}
\leavevmode
\epsfxsize=17cm \epsfbox{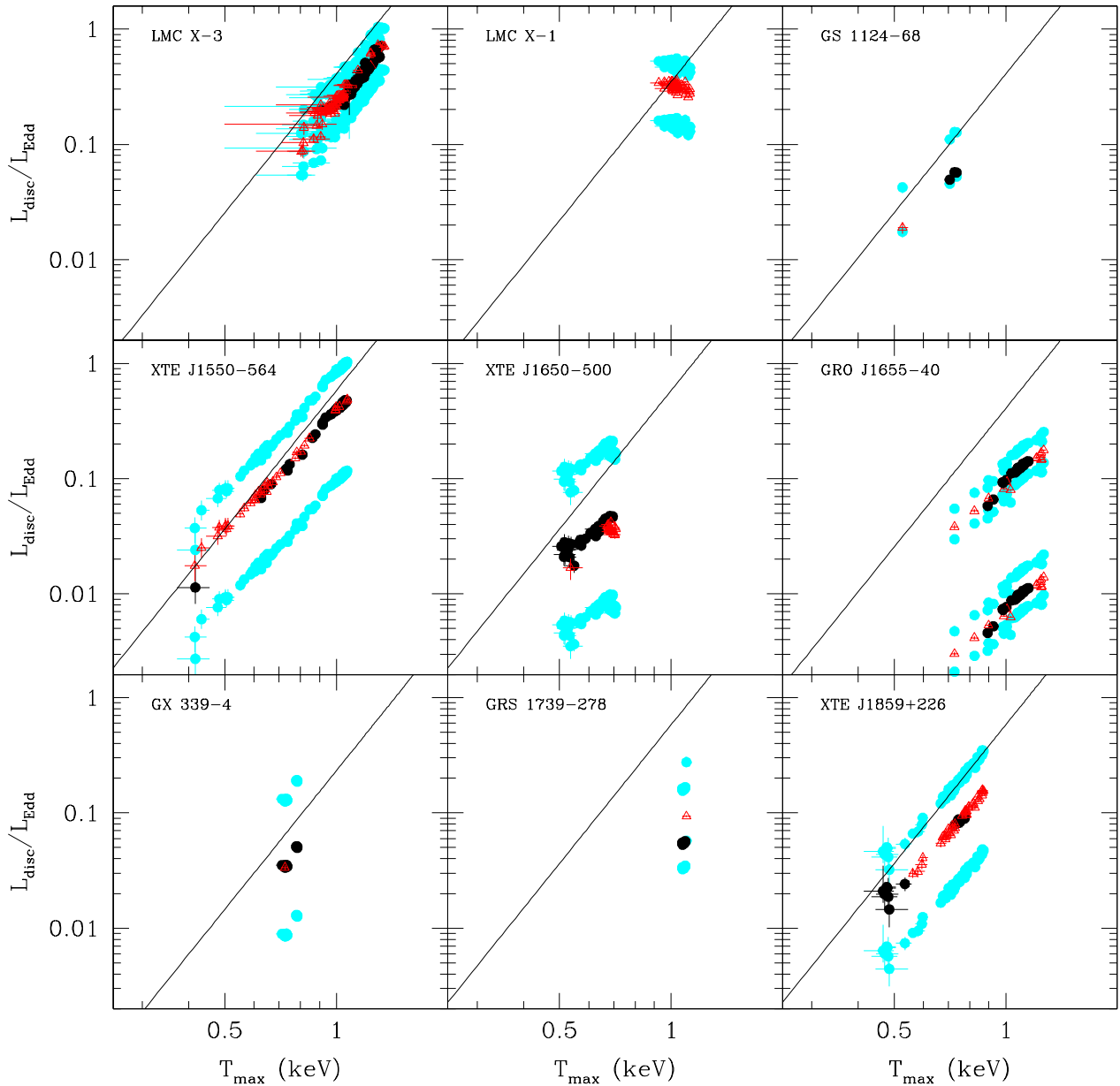}
\end{center}

\caption{Disc luminosity versus its maximum temperature in nine 
sources from Table \ref{tab:sources} (XTE~J2012+361 is shown 
separately in Fig.~\ref{fig:2012}). The black filled circles and red 
open triangles have the same meaning as in Fig.~\ref{fig:colum}. The 
cyan (or light grey in B\&W) points represent lower and upper limits 
on the data resulting from uncertainties in the distance and black 
hole mass (see Table \ref{tab:sources}). The lower and upper data 
points of GRO~J1655--40 correspond to distances of $0.9\pm0.1$ 
(Mirabel et al.~2002) and $3.2\pm0.2$ kpc (Hjellming \& Rupen 1995), 
respectively. The diagonal line plotted in each panel represents a 
relation $L_{\rm disc} \propto T_{\rm max}^4$ for a non-rotating 
black hole (Eq. \ref{eq:lt4}), calculated for the best-estimated mass 
(Table \ref{tab:sources}) and $f_{\rm col} = 1.8$. See electronic 
edition of the journal for a colour version of this figure.}

\label{fig:tlum}
\end{figure*}


\section{Results}
\label{sec:results}

\begin{figure}
\begin{center}
\leavevmode
\epsfxsize=7cm
\epsfbox{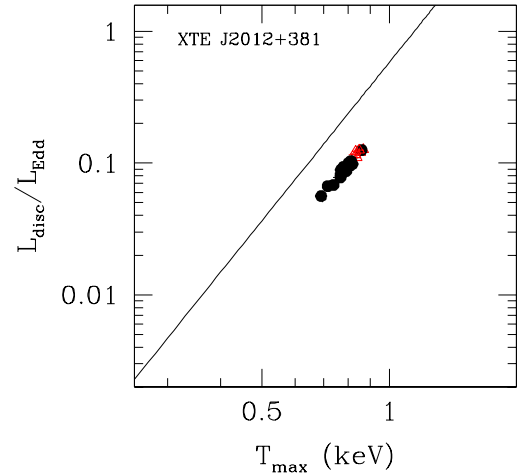}
\end{center}

\caption{Disc luminosity versus its maximum temperature for 
XTE~J2012+361. See Fig.~\ref{fig:tlum} for description of symbols 
and the diagonal line. There are no constraints for mass or 
distance for this source and we assume $M$ = 10M$_{\odot}$, $D$ 
= 10 kpc and $i = 60^\circ$.}

\label{fig:2012}
\end{figure}

The observed disc temperature-luminosity (hereafter $T$-$L$) 
relations are presented in Figs.~\ref{fig:tlum} and 
\ref{fig:2012}. They are 
roughly consistent with $L_{\rm disc} \propto T_{\rm 
max}^4$ for all the sources over a broad range of luminosity. 
This means there can be only little change in $f_{\rm col}$ with 
accretion rate. 

In each of the figures we also plot the expected 
$T$-$L$ relation from Eq.~(\ref{eq:lt4}) for a colour-correction 
factor of $f_{\rm col} = 1.8$. Obviously, the calculated $T_{\rm 
max}$ and $L_{\rm disc}/L_{\rm Edd}$ depend on rather uncertain 
values of the distance, mass and inclination angle (and unknown black 
hole spin). The major uncertainties from distance and mass are 
included in the figures; the cyan (or light grey in B\&W) shadows 
above/below the data points correspond to the upper distance and 
lower mass/lower distance and upper mass estimates, respectively. For 
three sources: XTE~J1650--500, GRS~1739--278 and XTE~J1859+226, where 
the mass is unknown, we have assumed a `canonical' black hole mass of 
10 M$_\odot$ with error range of 5--12 M$_\odot$ following the mass 
distribution of known sources (Bailyn et al.~1998). In case of 
XTE~J2012+381 there are no mass or distance estimates whatsoever, so 
we plot its results in a separate figure (Fig.~\ref{fig:2012}) for 
representative values of $M$ = 10 M$_\odot$ and $D$ = 10 kpc.

We stress that though both $T_{\rm max}$ and $L_{\rm disc}/L_{\rm 
Edd}$ depend on uncertain parameters, for a given source its 
distance, mass, spin and inclination are {\em constant}. Therefore, 
the {\em shape\/} of the pattern it traces in the $T$-$L$ diagram is 
robust. In particular, most of the observed sources 
follow a straight line in the $T$-$L$ diagram (which is a log-log 
plot) corresponding to a $L_{\rm disc} \propto T_{\rm max}^4$ 
relation. This can be particularly clearly seen for LMC~X-3 (see also 
Kubota et al.~2001), XTE~J1859+226 and XTE~J2012+381.

XTE~J1550--564 is particularly interesting, since it spans 
almost two orders of magnitude in luminosity. It generally 
follows $L_{\rm disc} \propto T_{\rm max}^4$ relation, though 
there is a small departure from this: with increasing 
temperature it seems to be slightly underluminous. This is 
particularly pronounced above $kT_{\rm max} \approx 0.9$ keV 
where a break in the $T$-$L$ relation can be seen. This has been 
also reported by Kubota \& Makishima (2003), who noticed that 
the data above the break is consistent with $L_{\rm disc} 
\propto T_{\rm max}^2$. They refer to this as the {\it 
apparently standard regime}, and suggest that it is associated 
with a transition to a slim disc (Abramowicz et al. 1988). 
However, this should only occur at/above $L_{\rm Edd}$ 
(Abramowicz et al.~1988; Shimura \& Manmoto 2003), so it seems 
more likely to represent a subtle change in colour temperature 
correction.

Other subtle deviations from a constant colour temperature correction
can also be seen. LMC~X-1 does not show much variability, but over the
observed small range it has an apparent anticorrelation of temperature
and luminosity. XTE~J1650--500 shows similar behaviour at the
high-temperature end of the $T$-$L$ diagram, where data points depart
from the $T_{\rm max}^4$ line.  In both cases the observations in
question show larger fraction of Comptonized luminosity. They might
belong to the {\it anomalous regime\/} (Kubota et al.~2001; Kubota \&
Makishima 2003), where Comptonization of the disc becomes
important. We note that the points on the low-temperature end of
XTE~J1650--500 have significant error bars and are consistent with $L_{\rm
disc} \propto T_{\rm max}^4$ relation.

\begin{figure}
\begin{center}
\leavevmode
\epsfxsize=8cm
\epsfbox{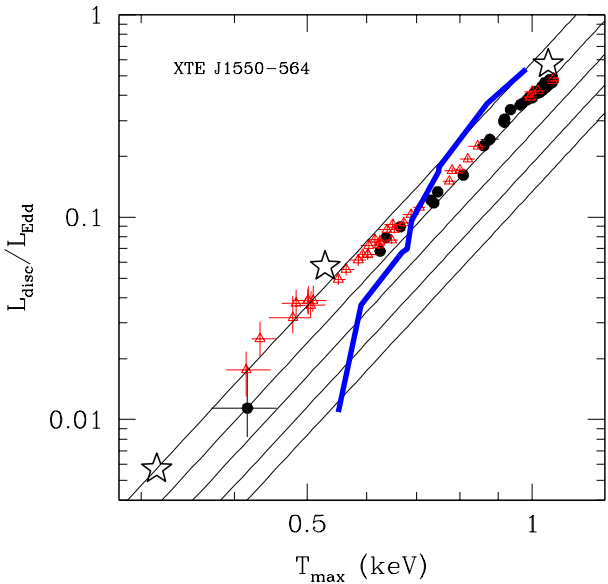}
\end{center}

\caption{Disc luminosity versus its maximum temperature for 
XTE~J1550--564 taken from Fig.~\ref{fig:tlum}. For clarity, we 
don't show distance/mass uncertainties, i.e. the data points 
here correspond to $D$ = 5.3 kpc and $M$ = 10 M$_\odot$. The 
diagonal lines are calculated for different values of the 
colour-correction factor, $f_{\rm col}$ = 1.8, 2.0, 2.2, 2.4 and 
2.6 (from top to bottom). The overplotted thick line is the 
expected $T$-$L$ relation from Merloni et al.~(2000), which 
predicts significant decrease in $f_{\rm col}$, with increasing 
accretion rate. Contrary to this prediction, the data are 
consistent with $f_{\rm col}$ remaining constant (or perhaps 
slightly increasing) over a very wide range of disc luminosity. 
The three stars represent calculations of Shimura \& Takahara 
(1995), which are in agreement with the data.}

\label{fig:merloni}
\end{figure}

The overall consistency of the disc-dominated
data with $L_{\rm disc} \propto T_{\rm max}^4$
relation means there is little change in $f_{\rm col}$ with luminosity
(see Eq.~\ref{eq:lt4}). The slight departures from this relation, seen
in GRO~J1655--40 and XTE~J1550--564 correspond to $f_{\rm col}$
increasing with luminosity. This is an important observational
constraint on disc models.  Shimura \& Takahara (1995) and Merloni et
al.~(2000) both calculate the colour temperature correction expected
from a standard SS73 disc around a black hole, including radiative
transfer through the vertical structure of the disc. Their predictions
are shown in Fig.~\ref{fig:merloni}, together with the XTE~J1550--564
data. The observations are clearly inconsistent with the Merloni et
al.~(2000) results, which predict a significant decrease of $f_{\rm
col}$ with luminosity. This shows that the vertical structure is {\em
not} as assumed in these calculations. On the other hand, these data
{\em are\/} consistent with predictions of Shimura \& Takahara (1995),
which are marked by stars in Fig.~\ref{fig:merloni}.

The absolute value of $f_{\rm col}$ (as opposed to its rate of change 
with $L_{\rm disc}/L_{\rm Edd}$) is somewhat more complex to 
constrain. At a given $L_{\rm disc}/L_{\rm Edd}$ the temperature can 
be higher if the black hole is spinning, or if there is continuous 
stress at the last stable orbit as opposed to the stress-free 
boundary condition. However, none of these effects will push the 
temperature {\em down}. For the best constrained source, GS~1124--68, 
the data show that the maximum value of the colour temperature 
correction is $\sim$~2.3. For other sources it can be much higher. 
There is no stringent lower limit on $f_{\rm col}$ from the data, as 
the same $T$-$L$ relation can be characterized be a black hole with 
higher spin and smaller $f_{\rm col}$.  Therefore, a conservative 
minimum value of $f_{\rm col}$ is unity, i.e. we see a true blackbody 
spectrum. However, this seems entirely unlikely given the high 
temperature and associated low absorption opacity. The lowest $f_{\rm 
col}$ suggested in this case is $\sim 1.6$ (Shimura \& Takahara 
1995). Considering GS~1124--68 to be a representative source, we can 
roughly limit $f_{\rm col}$ to be between 1.6 and 2.3.

Thus the use of $f_{\rm col}=1.8$ seems to be well justified. Many of 
the sources in Fig.~\ref{fig:tlum} lie close to the $T$-$L$ line for 
a Schwarzchild black hole with stress-free inner boundary condition. 
Again, the well constrained objects give the most information. A 
colour temperature correction of 1.8 implies that LMC~X-3 and 
GS~1124--68 are neither spinning {\em nor} have large amounts of 
energy released via continuous stress across the last stable orbit. 
Zero spin is clearly not a problem, but the lack of stress at the 
inner boundary is in conflict with the numerical simulations of 
(non-radiative) disc structure (Agol \& Krolik 2000).

\begin{figure*}
\begin{center}
\leavevmode
\epsfxsize=11.2cm \epsfbox{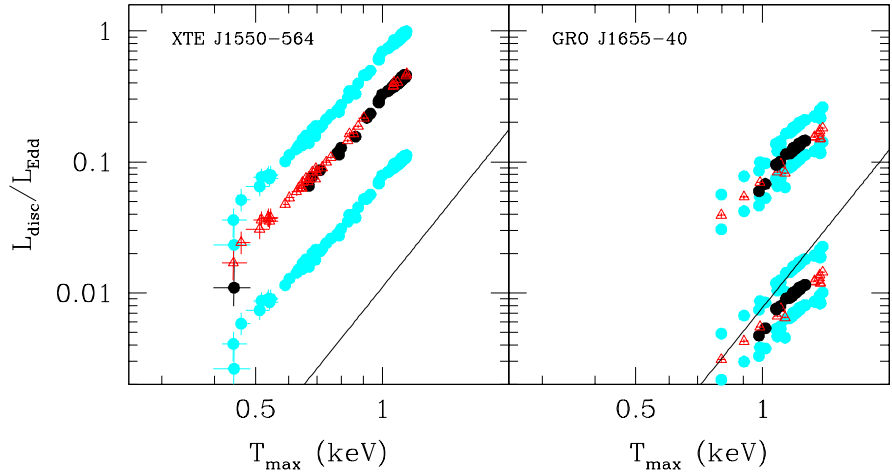}
\end{center}

\caption{Temperature-luminosity diagram for two sources selected from 
Fig.~\ref{fig:tlum}, but for a maximally rotating black hole ($a_* = 
0.998$). In this figure the diagonal line was calculated using 
formulae from Page \& Thorne (1974).}

\label{fig:tlum_kerr}
\end{figure*}

Several of the sources lie below the expected $T$-$L$ line, most 
noticeably GRO~J1655--40 (especially when $D$ = 0.9 kpc is assumed) 
and GRS~1739--278. Since we have already constrained the colour 
temperature correction and stress at the inner boundary then the only 
parameter left to change is spin. A spinning black hole might also 
have a different colour temperature correction or boundary stress 
condition, but we assume these stay constant to demonstrate the 
effect of the black hole spin.  Since the position of the $T$-$L$ 
line predicted by Eq.~(\ref{eq:lt4}) depends on the mass of the 
source, this creates another source of uncertainty. Therefore, we 
look at two sources with relatively precise mass estimates, namely 
XTE~J1550--564 and GRO~J1655--40. In Fig.~\ref{fig:tlum_kerr} we plot 
their $T$-$L$ diagrams once again, but this time assuming a maximally 
rotating black hole ($a_* = 0.998$, with the $T$-$L$ relation  
taken from Page \& Thorne 1974). Clearly, XTE~J1550--564 is not 
consistent with a maximal spin, while GRO~J1655--40 can be maximally 
rotating if its distance is given by the most recent determination 
of 0.9 kpc (Mirabel et al.~2002).

All the accretion discs observed here are apparently stable and 
show no sign of the limit-cycle variability expected at high 
accretion rates. To confirm this, we examine the light curves 
(16-s resolution) corresponding to analysed spectra. They are 
typically very smooth, with rms variability not exceeding 4 per 
cent. Only in the highest coronal fraction observations of 
LMC~X-1 and GRO~J1655--40 can the rms can reach 8--10 per cent. 


\section{Theoretical Colour Temperature Corrections}
\label{sec:fcol}

Fig.~\ref{fig:tlum} shows that in general the disc-dominated 
spectra have $L_{\rm disc}\propto T_{\rm col}^{1/4}$. This 
implies a {\em constant} colour temperature correction over a 
wide range of luminosity. This is somewhat unexpected from 
theoretical models of disc structure, as these predict that the 
emitted spectrum is formed under rather different conditions for 
different accretion rates, $\dot{m}_{\rm disc}$ ($\equiv \eta 
\dot{M}_{\rm disc} c^2/L_{\rm Edd}$, where accrestion efficiency 
$\eta = 1/16$). The first regime is at low $\dot{m}_{\rm disc}$ 
where the opacity is dominated by absorption processes and 
scattering is unimportant. The spectrum from a single radius is 
then very close to a simple blackbody of temperature $T_{\rm 
eff}$, so $f_{\rm col}=1$. What we see should be close to the 
disc blackbody predictions.

At higher $\dot{m}_{\rm disc}$ the temperature rises and the
absorption opacity drops, so scattering can become important. The
spectrum is distorted from a true blackbody to a modified blackbody
(e.g. Rybicki \& Lightman 1979). This has a lower emissivity than a
blackbody, but must still dissipate the same amount of gravitational
energy, so its `temperature' needs to be higher. Since the modified
disc blackbody spectrum depends on the absorption opacity, it is also
dependent on the density structure of the plasma. The resultant
spectrum differs depending on whether the density structure is
constant with height (radiation pressure dominated) or Gaussian (gas
pressure dominated: see equations 3.6 and 3.7 and fig.~2 of
SS73). However, there is also a further effect which is that radiation
can escape from much deeper into the disc, where the material is
hotter. Both these effects give an observed temperature which is
higher than $T_{\rm eff}$, so $f_{\rm col}>1$.

At even higher $\dot{m}_{\rm disc}$, the absorption opacity is so
small that scattering dominates. Emission from deep in the disc (very
hot material) can escape to the surface without being absorbed, but it
is downscattered by its many interactions with electrons in the cooler
outer layers of the disc. The spectrum is dominated by a Wien peak,
again giving $f_{\rm col}>1$ (see fig. 2 of SS73).

Merloni et al.~(2000) and Shimura \& Takahara (1995) show detailed
calculations of the spectra from standard $\alpha$-viscosity accretion
discs. Both find that Compton down-scattering and the vertical
temperature structure give rise to $f_{\rm col}\sim 1.8$ at high
$\dot{m}_{\rm disc} \gg 0.1$, where the disc is dominated by radiation
pressure.  However, their calculations differ at low $\dot{m}_{\rm disc}
\ll 0.1$ where the disc becomes gas pressure dominated (see
Fig.~\ref{fig:merloni}).  Both Merloni et al.~(2000) and Shimura \&
Takahara (1995) agree that Compton down-scattering is not important in
these spectra, but disagree on the resulting colour temperature
correction, giving $f_{\rm col}$ = 2.7 and 1.8, respectively.

There are potential problems with both sets of calculations. Shimura 
\& Takahara (1995) do not include heavy element opacity, so 
underestimate the true absorption in the disc, especially at 
low $\dot{m}_{\rm disc}$. Merloni et al.~(2000) do include these 
elements, but they assume that the vertical density profile is always 
given by that of a radiation pressure dominated disc i.e. is constant 
with height (SS73; Ross \& Fabian 1996). However, their lowest 
$\dot{m}_{\rm disc}$ calculations, where the colour temperature 
correction becomes very large, give a gas pressure dominated disc. 
Given that these calculations are in the modified disc blackbody 
regime, this would overpredict the temperature as the constant 
density discs give a larger colour temperature correction than the 
gas pressure dominated ones (fig.~2 of SS73). The extent of this 
overestimate can be estimated for an isothermal atmosphere from 
equations 3.6 and 3.7 of SS73 as $\sim 1.66/1.2=1.38$, which reduces 
the Merloni et al.~(2000) prediction to $f_{col}\sim 1.9$.

Thus with self-consistent vertical density structure, it appears that 
{\em both} Merloni et al.~(2000) and Shimura \& Takahara (1995) 
predict that the spectral hardening factor for Shakura-Sunyaev discs 
is more or less constant over the range of $\dot{m}_{\rm disc}$ 
explored here, as observed.

\section{Disc Stability}

Standard $\alpha$-discs are viscously and thermally unstable when 
radiation pressure, $P_{\rm rad}$, dominates over gas pressure, 
$P_{\rm gas}$. This is because the $\alpha$-viscosity prescription 
assumes that the viscous heating rate is proportional to the total 
pressure $P_{\rm tot}=P_{\rm rad}+P_{\rm gas}$.  A small increase in 
mass accretion rate produces a small rise in temperature, which leads 
to a much larger increase in heating in the radiation pressure 
dominated regime (as $P_{\rm rad}\propto T^4$) than in the gas 
pressure dominated case (where $P_{\rm gas}\propto T$). Radiative 
cooling cannot keep pace with the intense heating in the radiation 
pressure dominated disc, so these $\alpha$-discs are unstable at this 
point. The runaway heating can be halted when the $\alpha$-disc 
equations are extended to include radial energy transport (optically 
thick advection).  The advective cooling can balance even the intense 
$P_{\rm rad}$ heating, so these {\em slim} discs are stable 
(Abramowicz et al.~1988). 

However, while the $\dot{m}_{\rm disc}$ at which the disc becomes 
radiation pressure dominated is well defined, this merely marks the 
point at which the disc becomes {\em locally} unstable at a given 
radius. To translate this into {\em global} limit cycle of the inner 
disc requires that the instability can propagate to neighbouring 
radii, and this depends on details of the disc structure equations. 
Global numerical simulations with simple vertical disc structure show 
that this leads to limit cycle behaviour at $\dot{m}_{\rm disc}=0.06$ 
(Honma et al.~1991; Szuszkiewicz \& Miller 1997; 1998; Zampieri et 
al.~2001), while a more sophisticated vertical structure treatment 
shows that the disc is {\em globally} (but not locally) stable up to 
$\dot{m}_{\rm disc}=0.28$ (Janiuk, Czerny \& Siemiginowska 2002). 

The objects observed here span the range $0.01 \la L/L_{\rm 
Edd}\la 0.5$ and {\em all} of them are stable. The unusual 
microquasar GRS~1915+105 (not included in our sample) is the 
{\em only\/} source which shows variability which can be 
interpreted as this limit-cycle (Belloni et al.~1997). One 
possible explanation for this unique behaviour might be that 
GRS~1915+105 goes to higher $L/L_{\rm Edd}$ than the sources 
considered here, i.e. that the Shakura-Sunyaev disc models are 
indeed correct in predicting that there should be a radiation 
pressure instability, but that they underestimate the accretion 
rate required for the onset of the limit-cycle.

Plainly, the disc structure is different to that predicted by 
the standard $\alpha$-prescription. Either there are additional 
cooling mechanisms which act to stabilize the intense viscous 
heating predicted by these models, and/or the 
$\alpha$-prescription for viscous heating is incorrect. 

\subsection{Additional Cooling Mechanisms} 

The disc is cooler if some fraction, $f$, of the gravitational 
potential energy released by viscous heating is dissipated outside of 
the optically thick disc. This could also be associated with a mass 
accretion rate, $\dot{m}_{\rm loss}$, so that the total mass 
accretion rate is $\dot{m}=\dot{m}_{\rm disc} + \dot{m}_{\rm loss}$. 

In the limit of $\dot{m}_{\rm loss} = 0$ the disc structure can be 
derived under the standard $\alpha$-viscosity prescription (Svensson 
\& Zdziarski 1994). The disc is less luminous and cooler for a given 
$\dot{m}_{\rm disc}$, or alternatively, for a given disc luminosity 
and temperature the disc is denser, delaying the onset of the 
radiation pressure instability (Svensson \& Zdziarski 1994). The 
obvious candidate for this energy loss channel is in powering a corona 
above the disc, since hard X-ray emission is generally seen from 
these systems. However, we have chosen to use only spectra which have 
$f \la 0.15$, where Svensson \& Zdziarski (1994) show that the 
stability of the disc is very little changed from that of the 
straightforward SS73 disc. These models do not include irradiation of 
the disc by this corona, but again the limit of 15 per cent of the 
power dissipated in the hard X-ray emission means that it has very 
small effect on the disc structure (e.g. Nayakshin et al.~2000a).

Another candidate energy loss mechanism is to power an outflowing wind
or jet. Jets are commonly seen from black hole binaries {\em except}
when their spectra are disc dominated (e.g. Fender \& Kullkers 2001),
ruling this out as a major energy loss route. Winds could be present,
but line driving mechanisms which can power most of the mass loss
inferred from the UV emitting discs around supermassive black holes
and white dwarfs are {\em not} efficient in the much higher
temperature X-ray binary discs (Proga \& Kallman 2002).  A
magnetically driven wind is the only remaining possibility, in which
case the results depend on the form of the wind (e.g. Nayakshin,
Rappaport \& Melia 2000b). However, such a wind would have to be
unfeasibly powerful in order to stabilise the disc. Svensson \&
Zdzairski show that more than 95 per cent of the energy must be
dissipated in an invisible energy loss channel in order to produce a
stable disc at radiating at $L_{\rm disc}/L_{\rm Edd}=0.5$.

Models in which the energy dissipation is also associated with some 
fraction of the mass accretion are no better.  In the limit 
where the $f=\dot{m}_{\rm loss}/\dot{m}_{\rm disc}$, i.e. the energy 
released in the disc and alternative loss channel are proportional to 
their mass accretion rates, the disc is identical to a standard 
Shakura-Sunyaev disc with accretion rate $\dot{m}_{\rm disc}$ i.e. is 
unstable over the observed $0.01\la L/L_{Edd}\la 0.5$, and there is 
no obvious identification of this proposed energy/mass loss channel. 
A hard X-ray corona is still ruled out from lack of hard X-ray 
emission, jets are still ruled out from lack of radio emission, and 
(unobserved) winds would still need to be very powerful in order to 
stabilise the disc.  However, these models also include the energy 
exchange (via conduction and radiation) between the two accreting 
phases in the case of a corona. These `accreting corona' models 
calculate the vertical structure associated with an standard 
$\alpha$-disc, and show that a hot skin can develop over the main 
body of the disc. However, this corona is always optically thin, and 
becomes negligible at the high $L/L_{\rm Edd}$ considered here (Witt, 
Czerny \& {\.Z}ycki 1997; R{\'o}{\.z}a{\'n}ska \& Czerny 2000), so 
has no effect on the disc stability. 

\subsection{Alternative viscosity prescriptions}

The difficulties of avoiding the radiation pressure instability with
standard $\alpha$-discs make alternative prescriptions attractive.
One old idea is to replace the $\alpha$-viscosity (in which the
heating is proportional to the total pressure, $P_{\rm tot}=P_{\rm
gas}+P_{\rm rad}$) with one which simply depends only on $P_{\rm
gas}$. Such discs (hereafter termed $\beta$-discs) are stable even
when dominated by radiation pressure (Lightman \& Eardley 1974;
Sakimoto \& Coroniti 1981; Stella \& Rosner 1984), since the heating
is always $\propto T$. 

While such models trivially solve the disc stability, they give a 
very different disc structure and hence predict different behaviour 
for colour temperature correction as a function of accretion rate. 
For low $\dot{m}_{\rm disc}$, where gas pressure dominates, then the 
$\beta$- and $\alpha$-disc structures are identical, hence should 
have $f_{\rm col}\sim 1.8$ (see Section \ref{sec:fcol}). However, at 
high $\dot{m}_{\rm disc}$, where radiation pressure dominates, the 
density of the $\beta$-disc is much larger than that of the standard 
$\alpha$-disc (Stella \& Rosner 1984), though the disc thickness is 
unchanged (Stella \& Rosner 1984). Compton down-scattering is not 
important in these spectra at any $\dot{m}_{\rm disc}$ (Nannurelli \& 
Stella 1989), so the disc spectrum is in the modified blackbody 
regime. Its shape at any radius depends on how the density and 
temperature change as a function of height in the disc (SS73).  While 
full calculations have yet to be done, we can use the $\alpha$-disc 
calculations in a comparable regime (radiation pressure dominated, 
but where the emission is not strongly Comptonized) to estimate the 
resulting colour temperature corrections. This corresponds to the 
lowest $\dot{m}_{\rm disc}$ of the Merloni et al.~(2000) 
calculations, i.e. to $f_{\rm col}\sim 2.7$. Thus it seems that the 
$\beta$-disc viscosity predicts $f_{\rm col}$ increasing from $\sim 
1.8$ at low $\dot{m}_{\rm disc}$ to $\sim 2.7$ at high $\dot{m}_{\rm 
disc}$. This is strongly at odds with what is observed, leading us to 
conclude that the $\alpha P_{\rm gas}$ heating law is ruled out 
observationally (see also Janiuk et al.~2001).

Thus {\em none} of the {\it ad hoc} viscosity prescriptions appears
able to give a disc structure which matches both the observed colour
temperature corrections and stability. The answer probably lies in the
next generation of accretion disc models which will superseded these
{\it ad hoc} viscosity prescriptions. The physical mechanism for
angular momentum transport is now known to be the magneto-rotational
instability (MRI: Balbus \& Hawley 1991).  Numerical MHD simulations
of the global disc structure of gas pressure dominated discs show that
the $\alpha$-prescription applies approximately in the disc body (with
vertically integrated $\langle\alpha\rangle=0.1$). However, these also
show strong gradients in the ratio of viscous heating to pressure in
both radial and vertical directions, and large fluctuations in the
heating as a function of time at any point (e.g. Hawley \& Balbus
2002). For the bright discs considered here, radiation pressure
strongly affects the dynamics, and the mechanics of the MRI induced
turbulence changes (Agol \& Krolik 1998; Blaes \& Socrates 2001). The
coupled radiation and magneto-hydrodynamics is a difficult and very
computationally intensive problem, but preliminary results indicate
that the average heating is unlikely to be proportional to total
pressure, as in the standard $\alpha$-disc (Turner, Stone \& Sano
2002; Turner et al.~2003). Even if it were, the MRI induced turbulence
and associated convection may lead to enhanced cooling rates (Turner
et al.~2002; 2003)

Self-consistent disc models open up the possibility of being able to
{\em calculate} the viscous heating and convective turbulent cooling
associated with the MRI. When these models are sufficiently robust to
predict the vertical disc structure and associated radiation spectra,
then the data presented here offer a crucial test.

\section{Black Hole Spin}

`Black holes have no hair': they should be purely characterised by 
their mass and spin (any charge would always quickly be neutralized 
by accretion). The stellar remnant black holes all have very similar 
masses, so the only free parameter to give observable differences at 
the same (steady state) accretion rate is spin.  Prograde spin drags 
the last stable orbit closer to the black hole, resulting in higher 
luminosity and temperature for a given accretion rate onto a spinning 
black hole compared to a Schwarzschild one. This shifts the expected 
$T$-$L$ relation to the right (Fig.~\ref{fig:tlum}), giving an 
observable determination of the black hole spin (e.g. Zhang et 
al.~1997; Gierli{\'n}ski et al.~2001). Retrograde spin has the 
opposite effect, but is highly unlikely since the binary stars formed 
together from a single cloud, so spin and orbital angular momentum 
should be coupled.

Plainly, most  of the black  holes  in this  sample are consistent 
with Schwarzchild, lying on or close to the zero-spin $T$-$L$ 
relation. LMC~X-1, LMC~X-3 and  GS~1124--68, where the system 
parameters are tightly constrained, have very little room for even 
moderate spin. Conversely GRO~J1655--40 and GRS~1739--278 have 
significant spin. 

All these (apart from GRS~1739--278) were known previously: 
GS~1124--68 (Cui et al.~1997), GRO~J1655--40 (Cui et al.~1997; 
Gierli{\'n}ski et al.~2001; Kubota et al.~2001), LMC~X-3 (Cui et 
al.~1997; Kubota et al.~2001), LMC~X-1 (Gierli{\'n}ski et al.~2001). 
What is new here is the size of the sample, which shows clearly that 
a constant colour temperature correction is a fairly good 
approximation.  The rest of the objects have rather large 
uncertainties, and are compatible with moderate (but not extreme) 
spin. However, the fact that the best estimates for the system 
parameters put XTE~J1550--564, GX~339--4, XTE~J1650--500 and XTE~ 
J1859+228 so close to the Schwarzchild line is certainly suggestive, 
if not compelling. 

Better system parameters from optical/infrared studies of the systems 
will reduce the uncertainties, and provide a clear determination of 
the black hole spin. This is especially important (but for different 
reasons) for two of the systems here.  XTE~J1550--564 has a 
superluminal radio jet (Hannikainen et al.~2001; Corbel et al.~2002), 
so confirmation of its Schwarzchild nature would provide a conclusive 
counterexample to the paradigm in which relativistic jets are 
triggered by black hole spin (e.g. Moderski, Sikora \& Lasota 1998). 
XTE~J1650--500 is claimed to have maximal spin from detailed fitting 
of the iron line profile in the X-ray spectrum (Miller et al.~2002). 
Currently, with no good mass estimate our data cannot constrain its 
spin, though in future it might give an independent measurement. Here 
the issues are even more emotive, as the line profile in 
XTE~J1650--500 is similar to that of the AGN MCG--6--30-15 (Wilms et 
al.~2001), the defining example of a spinning black hole. We stress 
that the ultrasoft X-ray binary spectra offer an independent 
constraint on the spin, and hence give a check on iron line 
modelling. 

\section{Conclusions}

We observe discs emitting in the luminosity range $0.01 \la L_{\rm 
disc}/L_{\rm Edd} \la 0.5$. Individual objects which span a large 
range in $L_{\rm disc}/L_{\rm Edd}$ are generally consistent with 
$L_{\rm disc}\propto T^4$, implying a constant colour temperature 
correction. In LMC~X-3 and GS~1124--68, the distance, mass and 
inclination of the black hole are well constrained, giving $f_{\rm 
col} \sim 1.8$, and, by implication, require a Schwarzchild black 
hole. Conversely, for GRO~J1655--40 and GRS~1739--278 the disc 
temperature requires moderate-to-high spin, i.e. a Kerr geometry. For 
black hole binaries which show disc-dominated spectra this gives an 
alternative way to check the inner disc geometry derived from 
modelling the relativistic smearing of the iron line and reflected 
spectra. 

The observed range of $0.01 \la L_{\rm disc}/L_{\rm Edd} \la 0.5$ is 
where standard Shakura-Sunyaev models predict that the inner disc 
should be unstable due to the dominance of radiation pressure. The 
models predict limit-cycle behaviour in this range, yet observations 
show that the light curves are characterized by very little 
variability. These observations strongly require that the disc is 
stable in this range of $L_{\rm disc}/L_{\rm Edd}$.

Stability could be retrieved in the Shakura-Sunyaev disc models if 
the disc remains gas pressure dominated. This can be arranged if 
there are other cooling mechanisms (corona, jet, winds) which give a 
much denser disc for a given $\dot{m}_{\rm disc}$.  However, since we 
choose only disc-dominated spectra, the coronal emission is 
negligible. These spectra are also always observed to have strongly 
quenched radio emission, so limiting the jet emission, while winds 
would have to be unfeasibly powerful. Hence there is only very 
limited potential for avoiding the conclusion that radiation pressure 
does dominate in our high $L_{\rm disc}/L_{\rm Edd}$ discs.

The only other known way to circumvent the radiation pressure 
instability is to change the viscosity prescription so that the disc 
heating is proportional to $P_{\rm gas}$ rather than the standard 
$\alpha$-disc prescription where it is proportional to $P_{\rm 
tot}=P_{\rm gas}+P_{\rm rad}$. This avoids the tremendous increase in 
heating rate so the disc can remain stable even when dominated by 
radiation pressure. While proper calculations have yet to be made, it 
appears that such discs should have a colour temperature correction 
which changes $\sim$1.8 to $\sim$2.7 as $\dot{m}_{\rm disc}$ 
increases. Conversely, the standard $\alpha$-discs predict that the 
colour temperature correction should remain stable at $\sim$1.8 for 
the range of $\dot{m}_{\rm disc}$ considered here. The observation 
that the colour temperature correction remains stable over a wide 
range in $L_{\rm disc}/L_{\rm Edd}$ appears to rule out the 
alternative viscosity prescription, and strongly supports the 
vertical structure as predicted by a standard $\alpha$-disc.

This leads to an impasse. Observations show that the disc is most 
likely radiation pressure dominated, with standard 
$\alpha$-viscosity. Such discs should undergo limit cycle behaviour, 
yet are observed to be stable. There are strong indications that the 
solution to this lies in the self-consistent heating and cooling 
generated by the MRI turbulence which is the physical source of the 
viscosity. While coupled radiative magneto-hydrodynamic calculations 
cannot yet predict the disc vertical structure in a robust way, this 
will probably become possible in the next few years. 

Irrespective of the outcome of detailed calculations, it is clear 
that accretion disc spectra {\em can} constrain disc models.  While 
spectra which contain a substantial Compton component at higher 
energies are indeed ambiguous, we challenge the Merloni et al.~(2000) 
assertion that fitting the accretion disc spectrum gives results 
which are {\em not} in general directly related to the actual disc 
parameters. As also shown by Ebisawa et al.~(1994), Kubota et 
al.~(2001) and Kubota \& Makishima (2003), the disc dominated spectra 
{\em can be} a reliable guide to the accretion disc structure.

\section*{Acknowledgements}

We thank Omar Blaes, Bo{\.z}ena Czerny, Julian Krolik and Aya 
Kubota for useful discussions. This research has made use of 
data obtained from the High Energy Astrophysics Science Archive 
Research Center (HEASARC), provided by NASA's Goddard Space 
Flight Center.


\label{lastpage}

\end{document}